\documentclass[aps,prd,groupedaddress,showpacs,floatfix]{revtex4}
\usepackage{graphicx}
\usepackage{dcolumn}
\usepackage{bm}
\begin{document}

\title{Branching ratios of the decays of $\psi(3770)$ and $\Upsilon(10580)$ into light hadrons.}

\author{N.~N.~Achasov}
\email[]{achasov@math.nsc.ru}
\author{A.~A.~Kozhevnikov}
\email[]{kozhev@math.nsc.ru}
\altaffiliation{}
\affiliation{Laboratory of  Theoretical Physics, S.~L.~Sobolev Institute for Mathematics,
630090, Novosibirsk, Russian Federation}

\date{\today}

\begin{abstract}
Taking into account the new data on  the full width of
$D^{\ast\pm}(2010)$ and the mass difference of the charged and
neutral beauty mesons $B^\pm$,  $B^0,\bar B^0$, the branching
ratios of the decays $\psi(3770),\Upsilon(10580)\to\pi^+\pi^-$,
$K\bar K$, $\rho(\omega)\pi$,  $\rho(\omega)\eta$,
$\rho(\omega)\eta^\prime$, $K^\ast\bar K+{\rm c.c}$,
$\rho^+\rho^-$, and $K^\ast\bar K^\ast$ are re-evaluated in the
model in which the Okubo-Zweig-Iizuka rule is violated due to the
real intermediate state $D\bar D$ in case of $\psi(3770)$ and
$B\bar B$ in case of $\Upsilon(10580)$. The inclusive annihilation
of $\psi(3770)$ and $\Upsilon(10580)$ into light hadrons is
discussed.
\end{abstract}

\maketitle

\section{Introduction}

~ Long ago we  attracted  attention to the decays of heavy
quarkonia $\psi(3770)$ and $\Upsilon(10580)$ lying just above  the
corresponding open charm and beauty production thresholds, into
the states consisting of light mesons \cite{ach94}. Such decays
are interesting from the point of view of the study of the
two-step mechanism of the Okubo-Zweig-Iizuka (OZI) rule violation
\cite{lipkin87}. Recently interest in non-heavy-quark decays of
heavy vector quarkonia was renewed \cite{jon}.

As was argued in Ref.~\cite{ach94}, the small branching ratios of
the decays $J/\psi(1S)\to\omega\pi^0, \omega\eta,
\omega\eta^{\prime}, \rho\pi, \rho\eta, \rho \eta^{\prime},
K^{\ast} \bar K+{\rm c.c.} , \phi\eta, \phi\eta^{\prime}$ and
$\Upsilon (1S)\rightarrow\rho^0\pi^0, \pi^+\pi^-, K\bar K$ can be
understood in the framework of the dispersion approach as the
cancellation of the contributions of the intermediate states
$D\bar D$, $D^{\ast}\bar D+c.c.$, $D^{\ast}\bar D^{\ast}$ and so
on in the case of  $J/\psi(1S)$ [or $B\bar B$, $B^{\ast}\bar
B+c.c.$, $B^{\ast}\bar B^{\ast}$ and so on in the case of
$\Upsilon(1S)$]. In the meantime, there is no reason for large
suppression of the contribution of each specific intermediate
state. The cancellation can be violated when new channel opens.
Such a situation takes place, apparently, for the states
$\psi(3770)$ and $\Upsilon(10580)$ whose decay into light hadrons
proceeds via the real intermediate states  $D^+D^-+D^0\bar D^0$
and $ B^+B^-+B^0\bar B^0$, respectively. Hence, say, the
$\psi(3770)$ decay amplitude  to the pair of mesons $M_1$ and
$M_2$ is approximated by its imaginary part which can be obtained
from the unitarity condition,
\begin{eqnarray}
{\rm Im}M_{\psi(3770)\to D\bar D\to
M_1+M_2}&=&\frac{1}{2(2\pi)^2}\int\frac{d^3q_D}{2E_D}\frac{d^3q_{\bar
D}}{2E_{\bar D}}\delta^{(4)}(q_D+q_{\bar
D}-q_{M_1}-q_{M_2})\times\nonumber\\&&M^\ast_{D\bar D\to
M_1M_2}M_{\psi(3770)\to D\bar D},\label{unitarity}\end{eqnarray}
where
$$M_{\psi(3770)\to D\bar D}=g_{\psi(3770)D\bar
D}\epsilon_\mu(q_D-q_{\bar D})_\mu,$$ $\epsilon_\mu$ being the
polarization four-vector of the $\psi(3770)$, and $M_{D\bar D\to
M_1M_2}$ is the $D\bar D\to M_1M_2$ transition amplitude caused by
the meson exchange with the nonzero heavy flavor  quantum number.
The case of $\Upsilon(10580)$ is obtained from
Eq.~(\ref{unitarity}) by the evident replacements. The necessary
diagrams representing the amplitudes are given in
Ref.~\cite{ach94}.

 Since, first, the new crucial data on the full
width of $D^{\ast\pm}(2010)$  and on the mass difference of
charged and neutral $B$ mesons have appeared \cite{pdg}, second,
there are the plans to study such decays of the  $\psi(3770)$
meson with the upgraded CLEO-c facility \cite{asner}, and, third,
there is a huge number of the $\Upsilon(10580)$ mesons produced at
BABAR \cite{babar} and BELLE \cite{belle} that exceeds
$3\times10^8$, here, in Section \ref{exclus}, we re-evaluate the
branching fractions of the $\psi(3770)$ and $\Upsilon(10580)$
decay modes advertised above, taking into account the above
progress of the experiment. The inclusive annihilation of
$\psi(3770)$ and $\Upsilon(10580)$ into light hadrons via the
mechanism under discussion is considered in Section \ref{inclus}.
Note that such a consideration was absent in Ref. \cite{ach94}. In
addition, the Coulomb interaction effects between $D^+D^-$ and
$B^+B^-$ are discussed in Section \ref{coulomb}. Section
\ref{conclus} is devoted to a brief discussion of the background
problem and our hopes. Appendix contains some cumbersome technical
details of calculation of the $\psi(3770)$ and $\Upsilon(10580)$
decays into two vector states, which were not adduced in Ref.
\cite{ach94}.

\section{Exclusive decays}
\label{exclus}

 ~ First, we give the expressions for imaginary part
of the effective coupling constants of $\psi(3770)$ to the
$\pi^+\pi^-$ and $\omega\pi^0$ states. They can be obtained upon
using the following expressions for the amplitudes
$M_{D^+D^-\to\pi^+\pi^-}$ and $M_{D^+D^-\to\omega\pi^0}$:
\begin{equation}
M_{D^+D^-\to\pi^+\pi^-}=g^2_{D^\ast
D\pi^+}(q_{\pi^+}+q_{D^+},q_{\pi^-}+q_{D^-})\frac{\exp[\lambda_{D^\ast}(t-m^2_{D^{\ast0}})]}
{m^2_{D^{\ast0}}-t},\label{mddpp}\end{equation}
\begin{equation}
M_{D^+D^-\to\omega\pi^0}=2g_{D^\ast D\omega}g_{D^\ast
D\pi^0}\varepsilon_{\mu\nu\lambda\sigma}
(q_\omega)_\mu\omega_\nu(q_{\pi^0})_\lambda(q_{D^-})_\sigma
\frac{\exp[\lambda_{D^\ast}(t-m^2_{D^{\ast+}})]}
{m^2_{D^{\ast+}}-t},\label{mddvp}\end{equation} where
$\varepsilon_{\mu\nu\lambda\sigma}$ is totally antisymmetric unit
tensor, $\varepsilon_{0123}=-1$, $\omega_\nu$ is the $\omega$
meson polarization four vector,  the four-momentum $(q_a)_\lambda$
is labelled by the name $a$ of the corresponding particle,
$t=(q_{D^+}-q_{\pi^+})^2$, and $(a,b)=a_0b_0-\bm{a}\bm{b}$ denotes
the scalar product of the two four-vectors $a=(a_0,\bm{a})$ and
$b=(b_0,\bm{b})$. Note that the account is taken of the
possibility of the non-point-like $D^\ast$ exchange by means of
introducing the form factor whose exponential form is
parameterized by the slope $\lambda_{D^\ast}$ to be specified
below. The expressions for another isotopic state $D^0\bar D^0$
are obtained upon the evident replacements. The integration over
the two-particle intermediate states is reduced to the integration
over cosine of the angle $x=\cos\theta$ between the direction of
the momenta of the intermediate $D$ meson and the final meson
$M_1$:
\begin{eqnarray}
\frac{1}{(2\pi)^2}\int\frac{d^3p_1}{2E_1}
\frac{d^3p_2}{2E_2}\delta^{(4)}(q-p_1-p_2)M^\ast_{D\bar D\to
M_1M_2}M_{\psi(3770)\to D\bar D}=\frac{|{\bm
p}_1|}{8\pi s^{1/2}}\int d\cos\theta\times\nonumber\\
M^\ast_{D\bar D\to M_1M_2}M_{\psi(3770)\to D\bar
D}.\label{intint}\end{eqnarray} This results in the expressions
for imaginary parts of the required coupling constants:
\begin{eqnarray}
{\rm Im}g_{\psi(3770)\to\pi^+\pi^-}&=& \frac{g_{\psi
(3770)DD}g^2_{D^{\ast}D\pi^+}}{16\pi s^{1/2}q^2_{\pi\pi}}
\left\{q_{D^+D^-}\left[s-m^2_{D^+}-m^2_\pi+\frac{1}{2}
\left(m^2_{D^{\ast0}}+\right.\right.\right.\nonumber\\&&\left.\left.\left.
\frac{(m^2_{D^+}-m^2_\pi)^2}{m^2_{D^{\ast0}}}\right)\right]
\int_{-1}^1dx\frac{x\exp[2\lambda_{D^\ast}q_{D^+D^-}q_{\pi\pi}
(b_{D^+D^-}+x)]}{b_{D^+D^-}+x}\mp\right.\nonumber\\&&\left.
\mbox{contribution of } D^0\bar D^0\mbox{ intermediate
state}\right\}\approx\nonumber\\&&-4g^2_{D^\ast D\pi^+}
r_\mp\exp(-s\lambda_{D^\ast}/2),              
\label{gpsipp}\end{eqnarray}
\begin{equation}
b_{D^+D^-}=\frac{2(m^2_{D^+}+m^2_\pi-m^2_{D^{\ast 0}})-s}
{4q_{D^+D^-}q_{\pi\pi}},\label{b+-}\end{equation}
and
\begin{eqnarray}
{\rm Im}g_{\psi(3770)\to\omega\pi^0}&=&-\frac{g_{\psi
(3770)DD}g_{D^{\ast}D\omega}g_{D^{\ast}D\pi^0}} {8\pi
s^{1/2}q_{\omega\pi}}\left\{q^2_{D^+D^-}\int_{-1}^1dx\frac{1-x^2}
{a_{D^+D^-}+x}\times\right.\nonumber\\&&\left.
\exp[2\lambda_{D^\ast}q_{D^+D^-}q_{\omega\pi}(a_{D^+D^-}+x)]\mp
\right.\nonumber\\&&\left.\mbox{contribution of } D^0\bar
D^0\mbox{ intermediate state}\right\}\approx\nonumber\\&&
4g_{D^{\ast}D\omega}g_{D^{\ast}D\pi^0}r_{\mp}\exp(-s\lambda_{D^\ast}/2),\label{gpsivp}
\end{eqnarray}
\begin{equation}
a_{D^+D^-}=\frac{m^2_{D^+}+m^2_\omega-m^2_{D^{\ast
+}}-s^{1/2}E_\omega} {2q_{D^+D^-}q_{\omega\pi}}.
\label{a+-}\end{equation} Hereafter $s^{1/2}=m_{\psi(3770)}$
[$m_{\Upsilon(10580)}$] when discussing the decays of $\psi(3770)$
[$\Upsilon(10580)$], respectively. The contribution of the
$D^0\bar D^0$ intermediate state can be obtained from the $D^+D^-$
one in Eqs.~(\ref{gpsipp}), (\ref{b+-}), (\ref{gpsivp}), and
(\ref{a+-}) by the evident replacements. In the above formulas,
the approximate expressions containing $r_\mp$,  where
\begin{equation}
r_{\mp}=\frac{g_{\psi (3770)DD}}{6\pi m^3_{\psi (3770)}} \cdot
(q^3_{D^+D^-}\mp q^3_{D^0\bar D^0})                           
\label{r-+}\end{equation} are valid near threshold assuming the
approximate degeneracy of $D^\ast$ and $D$ mesons \cite{ach94}.
The sign plus (minus) corresponds to the situation when isospin is
conserved (violated) in the course of decay,
\begin{eqnarray}
E_b&=&\frac{1}{2m_a}(m^2_a+m^2_b-m^2_c),\nonumber\\
q_{bc}&=&\frac{1}{2m_a}\times\sqrt{\left[m^2_a-(m_b-m_c)^2]\times[m^2_a-(m_b+m_c)^2\right]}
\label{q_bc}\end{eqnarray}are, respectively, the center-of-mass
energy of the particle $b$, the momentum of the particle $b$ (or
$c$) in the final state of the decay $a\to b+c$ expressed through
the masses $m_{a,b,c}$. The necessary coupling constants are
evaluated as
\begin{equation}
|g_{\psi(3770)D\bar D}|=\left[\frac{6\pi
m^2_{\psi(3770)}\Gamma_{\psi(3770)\to D^+D^-+D^0\bar
D^0}}{q^3_{D^+D^-}+q^3_{D^0\bar D^0}}\right]^{1/2}=13.4,
\label{gpsidd}\end{equation}
\begin{equation}
|g_{D^\ast
D\pi^+}|=\left[\frac{6\pi\Gamma_{D^{\ast\pm}}\left(B_{D^{\ast\pm}\to
D^0\pi^\pm}+B_{D^{\ast\pm}\to D^\pm\pi^0}\right)}{q^3_{D^0\pi^\pm}
+\frac{1}{2}q^3_{D^\pm\pi^0}}\right]^{1/2}=9.1,\label{gdstdpi}\end{equation}
where $g_{D^\ast D\pi^0}=\frac{1}{\sqrt{2}}g_{D^\ast D\pi^+}$.
[Compare 9.1 in the right hand side of Eq.~(\ref{gdstdpi}) with
the figure of 4.5 resulting from the quark counting rule relation
$|g_{D^\ast D\pi^+}|\approx |g_{K^\ast K\pi^+}|$.] The coupling
constant $g_{D^{\ast}D\omega}$ cannot be found from existing data.
We choose the quark counting rule to obtain its magnitude:
\begin{equation}g_{D^{\ast}D\omega} \approx
g_{K^{\ast}K\omega}\approx \frac{1}{2}g_{\omega\rho\pi}=7.2\mbox{
GeV}^{-1},\label{coup1}\end{equation} where
$g_{\omega\rho\pi}=14.3\mbox{ GeV}^{-1}$ is obtained from the
branching ratio $B_{\omega\to\pi^+\pi^-\pi^0}$. Another relations
necessary for obtaining the rates of the decays $\psi
(3770)\to\omega\eta, \omega\eta^{\prime}, \rho\pi, \rho\eta,
\rho\eta^{\prime}$ from $B_{\psi(3770)\to\omega\pi^0}$ are
\begin{equation}
g_{D^{\ast}D\eta}\approx-\sqrt{\frac{2}{3}}g_{D^{\ast}D\pi^0}\approx
\sqrt{2}g_{D^{\ast}D\eta^{\prime}},\quad g_{D^{\ast 0}D^0\omega}=
-g_{D^{\ast
0}D^0\rho}=g_{D^{\ast+}D^-\rho},\label{coup2}\end{equation} where
the quark content $\eta=\sqrt{\frac{1}{3}}(u\bar u+d\bar d-s\bar
s)$, $\eta^\prime=\sqrt{\frac{1}{6}}(u\bar u+d\bar d+2s\bar s)$
corresponding to the pseudoscalar mixing angle
$\theta_P=-19.5^\circ$ is understood. To obtain
Im$g_{\psi(3770)\to \bar K^0K^0}$ [Im$g_{\psi(3770)\to K^-K^+}$]
from Im$g_{\psi(3770)\to D\bar D\to\pi^+\pi^-}$, one should take
into account the unique intermediate state $D^+D^-$ ($D^0\bar
D^0$) and the $D^\ast_s$ exchange in the amplitude $D^+D^-\to \bar
K^0K^0$ ($D^0\bar D^0\to K^-K^+$),  and make the proper
replacement in the coupling constants. The necessary relation is
\begin{equation}
g_{D^{\ast+}_sD^+K^0}=g_{D^{\ast+}D^0\pi^+}.\label{coup3}\end{equation}
Im$g_{\psi (3770)\to D\bar D\to\bar K^{\ast}K}$ can be found from
Im$g_{\psi (3770)\to D\bar D\to\omega\pi^0}$ in the same manner as
in the above case of the pseudoscalar final state. The necessary
relations are \begin{equation}g_{D^{\ast}_sDK^{\ast}}\approx
\sqrt{2}g_{D^{\ast}D^{\ast}\pi^0}
\approx\sqrt{2}g_{K^{\ast}K^{\ast}\pi^0} \approx
\frac{g_{\omega\rho\pi}}{\sqrt{2}}.\label{coup4}\end{equation} The
partial widths of the decays to the pair of pseudoscalars $PP$ and
vector+pseudoscalar $V+P$ are
\begin{equation}
\Gamma_{\psi(3770)\to PP}=\frac{|g_{\psi(3770)PP}|^2}{6\pi
m^2_{\psi(3770)}}q^3_{PP}\mbox{, } \Gamma_{\psi(3770)\to
VP}=\frac{|g_{\psi(3770)VP}|^2}{12\pi}q^3_{VP}.\label{width}\end{equation}
The adopted approximation permits one to take $q_{PP}\approx
q_{VP}\approx\frac{1}{2}m_{\psi(3770)}$.

The amplitude of the decay  $\psi(3770)\to\rho^+\rho^-,
K^{\ast}\bar K^{\ast}$ contains four independent structures:
\begin{eqnarray}
M(V\to V_1V_2)&=&\frac{1}{2}g_1(\epsilon^{(V)},q_1-q_2)(\epsilon
^{(V_1)}, \epsilon^{(V_2)})+g_2(\epsilon^{(V_2)},q_1)(\epsilon
^{(V)},\epsilon^{(V_1)})+\nonumber\\&&g_3(\epsilon
^{(V_1)},q_2)(\epsilon^{(V)},\epsilon
^{(V_2)})+\nonumber\\&&\frac{1}{2}g_4(\epsilon^{(V)},q_1-q_2)
(\epsilon^{(V_1)},q_2)(\epsilon^{(V_2)}),q_1),
\label{mvv}\end{eqnarray}where $\epsilon^{(V_{1,2})},q_{1,2}$
stands for the polarization and momentum four-vectors of the
corresponding vector particle.  The partial width is
\begin{eqnarray}
\Gamma_{\psi (3770)\rightarrow\rho^+\rho^-}(s)&=&
\frac{q^3_{\rho\rho}}{24\pi s}\Biggl\{2\Biggl[|g_1|^2+(|g_2|^2+
|g_3|^2)\frac{s}{m^2_\rho}\Biggr]+  \nonumber\\
&&|g_1+(g_2-g_3)\frac{s^{1/2}}{m_\rho}
+Gq^2_{\rho\rho}|^2\Biggr\}, \label{psi2rho}\end{eqnarray} where
$s=m^2_{\psi(3770)}$,
\begin{equation}
G=\frac{1}{m^2_\rho}\left[2g_1+\frac{2s^{1/2}(g_2-g_3)}
{(s^{1/2}+2m_\rho )}+g_4s\right]. \label{G}\end{equation}The
derivation of the expressions for imaginary parts of the coupling
constants $g_{1,2,3}$ and $G$ is outlined in Appendix.  The
results for the $K^{\ast-} K^{\ast+}$ ($\bar K^{\ast0} K^{\ast0}$)
final state are obtained from the $\rho^+\rho^-$ one by taking the
unique $D^0\bar D^0$ ($D^+D^-$) intermediate state contribution to
the unitarity relation Eq.~(\ref{unitarity}) and allowing for the
exchanges $D_s$ and $D^\ast_s$ in the $D\bar D\to \bar K^\ast
K^\ast$ reaction, and by using the quark counting rule relations
$|g_{\bar K^{\ast0}D^+D^-_s}|=\sqrt{2}|g_{\rho^0DD}|$, $|g_{\bar
K^{\ast0}D^+D^{\ast-}_s}|=\sqrt{2}|g_{D^\ast D\rho^0}|$.

We use  $\lambda_D\approx\lambda_{D^\ast}=0.27$ GeV$^{-2}$ and
$\lambda_B\approx\lambda_{B^\ast}=0.04$ GeV$^{-2}$, that is, as
before in Ref.~\cite{ach94} we expect that
$\lambda_{D,D^\ast}\sim1/m^2_{D^\ast}$ and
$\lambda_{B,B^\ast}\sim1/m^2_{B^\ast}$. In addition, as  before in
Ref.~\cite{ach94}, we take into account the two-fold suppression
of the amplitudes due to the absorption in the final state. As a
result, the branching ratios are suppressed by factor 30 in
comparison with the ones calculated in the model with the point
like exchange. If to take courage for use the Regge exchange
\cite{slopes} at low energy, $\ln(s/s_0)\approx 1$, one get the
branching ratios nearly our ones. So we hope that our estimations
are conservative enough.

The resulting branching ratios, together with the expected number
of events adopted to the recent or planned CLEO integral
luminosity \cite{asner}, and the ratio of the signal to the
exclusive background estimated in the simple vector dominance
model are collected in Table \ref{tabpsi}. Note that the
absorption suppression factor 1/4 is excluded in the case of the
final states containing $J/\psi$ meson because both the
intermediate and final states in the reaction $D\bar D\to
J/\psi+\pi^0(\eta)$ are in the threshold region.
\begin{table}
\caption{Branching ratios of the $\psi(3770)$ decays into the pair
of light mesons. Also estimated  is  the expected number of events
$N_{1,2}$, where 1 and 2 correspond to the integral luminosity
$\int{\cal L}dt=60.3{\rm pb}^{-1}$  and $3{\rm fb}^{-1}$ attained
and planned at CLEO-c\cite{asner}. The quantities without (with)
parentheses correspond to the isospin $I=0$ taking place in the
model of $c\bar c$ quarkonium or $D\bar D$ molecule/four-quark
state with zero isospin ($D\bar D$ molecule/four-quark isovector
state). The charged and neutral pairs of strange particles have
coincident branching ratios for both values of isospin.}

\begin{center}
\begin{tabular}{lcccc}                \hline\hline
mode $f$& $B_{\psi (3770)\to f}$ & $N_1$&$N_2$ &signal/background \\
\hline $\pi^+\pi^-$ & $3\times 10^{-5}(7\times 10^{-4})$ & 20
(490)&1000 (24500)&0.1 (2)\\
$K^+K^-$ & $9\times 10^{-5}$ &60&3000&0.2 \\
$K^0\bar K^0$ & $9\times 10^{-5}$ & 60&3000&5 \\
$\omega\pi^0$ & $6\times 10^{-5} (2\times 10^{-3})$ & 40
(1400)&2000 (70000)&0.004 (0.1)
\\$\omega\eta$ & $1\times 10^{-3} (4\times 10^{-5})$ & 700 (30)&35000 (1550)&2 (0.1) \\
$\omega\eta^{\prime}$ & $5\times 10^{-4} (2\times 10^{-5})$ & 350 (10)&17500 (500)&2 (0.1)\\
$\rho\pi$ & $5\times 10^{-3} (2\times
10^{-4})$ &3500 (140)&175000 (7000)&1 (0.05)             \\
$\rho\eta$ & $4\times 10^{-5} (1\times 10^{-3})$ & 30 (700)&1500 (35000)&0.01 (0.4)\\
$\rho\eta^{\prime}$ & $2\times 10^{-5} (5\times 10^{-4})$ & 15
(350)& 750 (17500)&0.01 (0.4)\\
$K^{\ast+}K^-+{\rm c.c}$ & $8\times 10^{-4}$ & 560&28000&0.5\\
$K^{\ast0}\bar K^0+{\rm c.c}$ & $8\times 10^{-4}$ & 560&28000&0.05\\
$\rho^+\rho^-$ & $7\times 10^{-5} (2\times 10^{-3})$ &40
(1400)&2000 (70000)&0.003 (0.007) \\
$K^{\ast+}K^{\ast-}$ & $4\times 10^{-4}$ &270&14000&0.01\\
$K^{\ast0}\bar K^{\ast0}$ & $4\times 10^{-4}$ &270&14000&0.3\\
$J/\psi+\pi^0$&$2\times10^{-5}$ ($5\times10^{-4}$)&15 (350)&750 (17500)&$-$\\
$J/\psi+\eta$&$8\times10^{-5}$ ($3\times10^{-6}$)&60 (2)&3000 (100)&$-$\\
\hline\\
$\sum_fB_{\psi(3770)\to f}$& 0.009 (0.009)&$\sim6000$
($\sim6000$)&$\sim3\times10^5(\sim3\times10^5)$\\$B_{\psi(3770)\to{\rm3gluons}}$
& $2\times 10^{-4}$ & $-$&$-$
\\ \hline    \hline
\end{tabular}\end{center}\label{tabpsi}\end{table}

Almost all said above about $\psi (3770)$ can be translated to the
case of  $\Upsilon(10580)\equiv\Upsilon(4S)$ meson decays to light
mesons by means of the replacements $\psi(3770)\to\Upsilon(10580),
c\to b, D\to B, D^{\ast}_s\to B^{\ast}_s$ etc. The suppression due
to the vertex form factor and the absorption in the final state is
assumed to be the same as in the case of $\psi(3770)$. The
difference is that the electromagnetic mass difference of the
vector $B^{\ast+}$ and  $B^{\ast0}$ mesons is not known. Our
assumption is the above mass difference is vanishing. Using the
quark model relations expressing all unknown coupling constants
through the known ones which are analogous to those used in the
case of the $\psi(3770)$ decays, one can evaluate the branching
fractions of the $\Upsilon(10580)$ decays. The results of this
evaluation, again taking into account the still unruled
possibility of the isospin $I=1$ are given in the table
\ref{upsresults}. Note that the absorption suppression factor 1/4
is excluded in the case of the final states containing
$\Upsilon(1S)$ meson because both the intermediate and final
states in the reaction $B\bar B\to
\Upsilon(1S)+\pi^0(\eta,\eta^\prime)$ are in the threshold region.
Also given is  the ratio of the signal to the exclusive background
estimated in the simple vector dominance model.
\begin{table}
\caption{Branching ratios of the $\Upsilon(10580)$ decays into the
pair of light mesons. Also estimated  is  the expected number of
events $N$  corresponding to the integral luminosity $\int{\cal
L}dt=81.7{\rm fb}^{-1}$   attained at BABAR \cite{babar}. The
number of events expected in the whole statistics $\int{\cal
L}dt=140{\rm fb}^{-1}$ at BELLE \cite{belle} is obtained by
multiplying $N$ by the factor 1.7. The quantities without (with)
parentheses correspond to the isospin $I=0$ taking place in the
model of $b\bar b$ quarkonium or $B\bar B$ molecule/four-quark
state with zero isospin ($B\bar B$ molecule/four-quark isovector
state). The charged and neutral pairs of strange particles have
coincident branching ratios for both values of isospin.}
\begin{center}
\begin{tabular}{lccc}                \hline\hline
mode $f$ &$B_{\Upsilon(10580)\to f}$&$N$&signal/background\\\hline
$\pi^+\pi^-$&$7\times10^{-8}$ ($9\times10^{-5}$)&20 ($2.7\times10^4$)&0.03 (40)\\
$K^+K^-$&$2\times10^{-5}$&6000&6\\$K^0\bar
K^0$&$2\times10^{-5}$&6000&150\\$\omega\pi^0$&$1\times10^{-6}$
($1\times10^{-3}$)&300 ($3\times10^5$)&0.001 (2) \\
$\omega\eta$&$9\times10^{-4}$ ($7\times10^{-7}$)&$3\times10^5$
(240)&40 (0.03)
\\$\omega\eta^\prime$&$5\times10^{-4}$ ($4\times10^{-7}$)&$1.5\times10^5$
 (120)&40 (0.03)\\$\rho\pi$&$4\times10^{-3}$ ($3\times10^{-6}$)&
$1.5\times10^6$ (1200)& 30 (0.02)\\$\rho\eta$&$7\times10^{-7}$
($1\times10^{-3}$)&240 ($3\times10^5$)&0.006
(7)\\$\rho\eta^\prime$&$4\times10^{-7}$ ($5\times10^{-4}$)&120
($1.5\times10^5$)&0.006 (7)\\$K^{\ast+}K^-+{\rm
c.c}$&$1\times10^{-3}$&$3\times10^5$&18\\$K^{\ast0}\bar K^0+{\rm
c.c}$&$1\times10^{-3}$&$3\times10^5$&2\\
$\rho^+\rho^-$&$6\times10^{-6}$ ($7\times10^{-3}$)&1500
($2\times10^6$)&0.004
(7)\\
$K^{\ast+}K^{\ast-}$&$2\times10^{-3}$&
$6\times10^5$&1\\$K^{\ast0}\bar
K^{\ast0}$&$2\times10^{-3}$&$5\times10^5$&25\\
$\Upsilon(10580)+\pi^0$&$1\times10^{-7}$ ($1\times10^{-4}$)&30 (30000)&$-$\\
$\Upsilon(10580)+\eta$&$6\times10^{-5}$ ($4\times10^{-8}$)&18000 (10)&$-$\\
$\Upsilon(10580)+\eta^\prime$&$6\times10^{-6}$ ($5\times10^{-9}$)&1800 (1)&$-$\\
\hline\\
$\sum_fB_{\Upsilon(10580)\to f}$&0.012 (0.016)&$\sim4\times10^6$
($\sim5\times10^6$)&$-$\\
$B_{\Upsilon(10580)\to{\rm3gluons}}$&$6\times10^{-4}$&$-$&$-$
\\\hline\hline
\end{tabular}\end{center}\label{upsresults}\end{table}

\section{Inclusive decays}
\label{inclus}

~ Let us discuss the inclusive annihilation $\psi(3770)$ and
$\Upsilon(10580)$ into light hadrons, $\psi(3770)\to D^+D^- +
D^0\overline{D^0}\to X$ and $\Upsilon(10580)\to B^+B^- +
B^0\overline{B^0}\to X$. Supposing that the isotopic symmetry is
broken only by the mass difference of charge and neutral particles
in an isotopic multiplet and by electromagnetic interaction
between $D^+D^-$ and $B^+B^-$, we can express the total branching
ratios of the decays of $\psi(3770)$ and $\Upsilon(10580)$
  into light hadrons $X$
   via the  branching ratios
of the $\psi(3770)\to D\overline{D}$ and $\Upsilon(10580)\to
B\overline{B}$ decays and the annihilation cross-sections of the
$D\overline{D}$ and $B\overline{B}$
  real intermediate states into light hadrons $X$
 \begin{eqnarray}
 \label{inclusive}
 &&\sum_X B_{V\to P^+P^- + P^0\overline{P^0}\to X}=\left(m_V^2v_P^2/48\pi\right )\times B_{V\to
 P^0\overline{P^0}}\times\nonumber\\[9pt]
 && \sum_X\left [\,1+(-1)^{I_X+I_V}  |c_{P^\pm}|^2\times \left (v_{P^\pm}/v_{P^0}\right )^3\ \right
 ]^2\times\sigma_P\left (P^0\overline{P^0}\to X\right
 )\,,
 \end{eqnarray}
 where $V=\psi(3770)$ (or $\Upsilon(10580)$); $P (\overline{P}) = D (\overline{D})$
 (or $B (\overline{B})$); where $v_{P}$ is velocity of $P(\mbox{or}\ \overline{P})$ in c.m.
 system;
 $I_X$ and $I_V$ are
 the  isotopic spins of $X$ and $V$, respectively;
 $c_{P^\pm}$ takes into account electromagnetic interaction between $P^+$
 and $P^-$; $\sum_X\sigma_P\left (P^0\overline{P^0}\to X\right
 )$ is the total annihilation cross-section $P^0\overline{P^0}\to
 X$ in the $P$ wave with $I_X$=0, or 1. When deriving Eq. (\ref{inclusive}) terms of
order $\left  (v_{P^\pm}^n -v_{P^0}^n\right )$  are neglected in
\begin{equation}
\label{ratio}
 \sum_X\sigma_P\left (P^+P^-\to X\right
 ) /\sum_X\sigma_P\left (P^0\overline{P^0}\to X\right
 )\,,
\end{equation}
  where $n$=1, 2, ...; $v_{D^\pm}\approx 0.128$  and
 $v_{D^0}\approx 0.147$; $v_{B^\pm}\approx 0.064$  and
 $v_{B^0}\approx 0.063$.

  The factors of suppression of the foreign isotopic spin
  $(-1)^{I_X+I_V}=-1$ in Eq. (\ref{inclusive})
  \begin{equation}
  \label{suppressiond}
 d=\left [\,1-  |c_{D^\pm}|^2\times \left (v_{D^\pm}/v_{D^0}\right )^3\ \right ]^2/\left [\,1 +
|c_{D^\pm}|^2\times \left (v_{D^\pm}/v_{D^0} \right )^3\ \right
 ]^2\approx 0.04
\end{equation}
and
\begin{equation}
\label{suppressionb}
  b=\left [\,1-  |c_{B^\pm}|^2\times \left (v_{B^\pm}/v_{B^0}\right )^3\ \right
 ]^2/\left [\,1 +  |c_{B^\pm}|^2\times \left (v_{B^\pm}/v_{B^0}\right )^3\ \right
 ]^2\approx 0.0008
  \end{equation}
  at $c_{P^\pm}=1$.

  So, measuring the total branching ratios  of  the decays of $\psi(3770)$ and $\Upsilon(10580)$
  into light hadrons gives information about the $P$ wave annihilation cross-sections of $D\overline{D}$ and
  $B\overline{B}$ into light
  hadrons with the proper isotopic spin $(-1)^{I_X+I_V}=1$.
   At present experiment allows to the value of $\sum_X B_{\psi(3770)\to
  X}$ to be up to 10\% \cite{rosner}.
\begin{equation}
\label{percents}
   \sum_X B_{\psi(3770)\to D^+D^- + D^0\overline{D^0}\to
  X}=1\%\ \ \mbox{and}\ \ \sum_X B_{\Upsilon(10580)\to B^+B^- + B^0\overline{B^0}\to
  X}=1\%
  \end{equation}
    correspond
    \begin{equation}
    \label{crosssection}
    \sum_X\sigma_P(D^0\overline{D^0}\to X)\approx 1.5\,\mbox{mb\ \ and}\ \ \sum_X\sigma_P(B^0\overline{B^0}\to X)\approx 0.64\,\mbox{mb}\,,
    \end{equation}
  respectively.  Note that  $\sum_X\sigma_P(D^0\overline{D^0}\to X)\sim v_{D^0}$ and
 $\sum_X\sigma_P (B^0\overline{B^0}\to X)\sim v_{B^0}$.

\section{Coulomb corrections}
\label{coulomb}

~ Current experiment \cite{cleo,babar05} allows to draw a
conclusion about the electromagnetic corrections ( the
$|c_{P^\pm}|^2$ factors). CLEO gives
\begin{equation}
\label{cleo} r^{ex}_D=\sigma (e^+e^-\to D^+D^-)/\sigma (e^+e^-\to
D^0\overline{D^0})= 0.776 \pm 0.024 ^{+ 0.014}_{- 0.006}
\end{equation}
 at
$E_{cm}=3773$ MeV. The theoretical prediction at $c_{D^\pm}=1$ is
\begin{equation}
\label{rth} r^{th}_D= 0.6913 \pm 0.0085\,.
\end{equation} So, experiment \cite{cleo} gives
\begin{equation}
\label{cd} |c_{D^\pm}^{ex}|^2= r^{ex}_D/r^{th}_D = 1.123 \pm 0.033
^{+ 0.02}_{-0.01}\,,
\end{equation}
that does not contradict  the point like electromagnetic
correction \cite{1969}
\begin{equation}
\label{pointd} |c_{D^\pm}^{th}|^2\approx 1
+\frac{\alpha\pi}{2v_{D^\pm}}= 1.086 \pm 0.001\,.
\end{equation}
Note that Eq. (\ref{cd}) leads to the additional suppression in
Eq. (\ref{suppressiond}): $d\to 0.02$. As for $c_{B^\pm}$, one can
get an information about this  in indirect way from BABAR
\cite{babar05}
\begin{equation}
\label{babar}
 B_{\Upsilon(10580)\to B^0\overline{B^0}} = 0.487 \pm
0.010 (stat) \pm 0.008 (sys)\,.
\end{equation}
Supposing the two channel dominance $\Upsilon(10580)\to B^+B^-\,,\
B^0\overline{B^0}$, one can get the theoretical predictions
\begin{equation}
\label{1th} B_{\Upsilon(10580)\to B^0\overline{B^0}}^{1th} = 0.489
\pm 0.010
\end{equation}
at $c_{B^\pm}^{1th}=1$ and
\begin{equation}
\label{2th} B_{\Upsilon(10580)\to B^0\overline{B^0}}^{2th} = 0.448
\pm 0.010
\end{equation}
at the point like electromagnetic correction
\begin{equation}
\label{pointb} |c_{B^\pm}^{2th}|^2\approx 1
+\frac{\alpha\pi}{2v_{B^\pm}}= 1.178 \pm 0.004\,.
\end{equation}
So, there is good agreement between the data Eq. (\ref{babar}) and
the calculation without an electromagnetic correction Eq.
(\ref{1th}), at the same time there are three standard deviations
between the experiment value Eq. (\ref{babar}) and the calculation
taking into account the point like electromagnetic interaction Eq.
(\ref{2th}). This can be considered  as an evidence for the role
of the $B^+(B^-)$ structure \cite{lepage} ( or violation of
isotopic symmetry for the coupling constants
$g_{\Upsilon(10580)D^+D^-}$ and
$g_{\Upsilon(10580)D^0\overline{D^0}}$ ). The factor
$|c_{B^\pm}|^2 = 1.08$ would give $B_{\Upsilon(10580)\to
B^0\overline{B^0}} = 0.469$ consistent with experiment  within one
standard deviation. Note that $|c_{B^\pm}|^2 = 1.08$ leads to the
increase in Eq. (\ref{suppressionb}): $b\to 0.0044$. Whereas Eq.
(\ref{pointb}) would lead to the increase in Eq.
(\ref{suppressionb}): $b\to 0.0119$.

Note that the correction factors Eq.~(\ref{pointd}) and
(\ref{pointb}) result, respectively, in suppressing by the factor
of 0.6 of the $\psi(3770)$ decays  with foreign isospin Table
\ref{tabpsi}, which is beyond the accuracy of the calculation, and
in enhancing by the factor of 17 of the analogous processes of the
$\Upsilon(10580)$ decays in Table \ref{upsresults}. This follows
from the replacement of the factor $r_\mp$ Eq.~(\ref{r-+}) by the
modified factor
\begin{equation}
r^{\rm modified}_{\mp}=\frac{g_{\psi (3770)DD}}{6\pi m^3_{\psi
(3770)}} \cdot
(q^3_{D^+D^-}\times|c_{D^\pm}^{th}|^2\mp q^3_{D^0\bar D^0})                           
\label{r-+mod}\end{equation} and the analogous replacement in the
case of $\Upsilon(10580)$.

\section{Conclusion}
\label{conclus}

The inclusive background at $\psi(3770)$ is
\begin{equation}
\label{inclusivephonpsi} \sigma\left (e^+e^-\to u\bar{u} +
d\bar{d}+ s\bar{s}\,,\ \sqrt{s}=3770\,\mbox{MeV}\right )\approx
1.22\cdot 10^{-32}\,\mbox{cm}^2\,,
\end{equation}
the signal is
\begin{equation}
\label{inclusivesignalpsi} \sum_X\sigma\left (e^+e^-\to\psi
(3770)\to X\right )\approx 1.19\cdot B_{\psi(3770)\to X}\cdot
10^{-32}\,\mbox{cm}^2\,.
\end{equation}
The inclusive background at $\Upsilon(10580)$ is
\begin{equation}
\label{inclusivephonupsilon} \sigma\left (e^+e^-\to u\bar{u} +
d\bar{d}+ s\bar{s}\,,\ \sqrt{s}=10580\,\mbox{MeV}\right )\approx
1.55\cdot 10^{-33}\,\mbox{cm}^2\,,
\end{equation}
the signal is
\begin{equation}
\label{inclusivesignalupsilon} \sum_X\sigma\left
(e^+e^-\to\Upsilon (10580)\to X\right )\approx 3.77\cdot
B_{\Upsilon(10580)\to X}\cdot 10^{-33}\,\mbox{cm}^2\,.
\end{equation}
So, the relation signal/background is better in exclusive decays
in many cases, compare the tables \ref{tabpsi}  and
\ref{upsresults} with the ratios of Eq.~(\ref{inclusivesignalpsi})
to Eq.~(\ref{inclusivephonpsi}) and
Eq.~(\ref{inclusivesignalupsilon}) to
Eq.~(\ref{inclusivephonupsilon}).

The exclusive background is estimated as an incoherent one using
the simple vector dominance model, see the tables \ref{tabpsi} and
\ref{upsresults}. Certainly, the problem of the  background should
be discovered only in the way. Nevertheless it is clear even now
that the exclusive background is overvalued by a factor three in
the $\omega\pi^0$ mode and by a factor  ten in the $\rho^0\eta$
mode, see \cite{formfactorsinpsi}. Note that interference effects
can expand experimental possibilities.

So, the study of non-heavy-quark decays of heavy vector quarkonia
promises the exciting cruise unless discovery a new land
(four-quark nature of $\psi(3770)$ or $\Upsilon(10580)$, for
example). But in any case Foundation of Low Energy Physics of
Charmed and Beauty hadrons will be laid.

\begin{acknowledgments}
The present study was partially supported by the grants
RFFI-02-02-16061 from Russian Foundation for Basic Research and
NSh-2339.2003.2 for Support of Leading Scientific Schools. We
thank Misha Achasov, Karl Berkelman, and Jon Rosner for
discussions. N.N. Achasov thanks  David Cassel, Misha Dubrovin,
Hanna Mahlke-Krueger, Jon Rosner, Werner Sun, and Maury Tigner for
the generous hospitality in Cornell.
\end{acknowledgments}

\appendix*
\label{app}
\section{}
Here we outline the derivation of the contributions to the
imaginary parts of the coupling constants $g_{1,2,3}$ and $G$
standing in the $V\to V\bar V$ decay amplitude, where
$V=\psi(3770)$, $\Upsilon(10580)$, and $V=\rho^+,
K^{\ast+},K^{\ast0}$. See Eqs.~(\ref{mvv}) and (\ref{G}). To be
specific, let us choose the final state $\rho^+\rho^-$ to discuss
the details of the derivation. There are two contributions to the
amplitude of the reaction $D^+D^-\to\rho^+\rho^-$ due to the $D$
and $D^\ast$ exchanges. They are:
\begin{eqnarray}
M^{(D)}_{D^+D^-\to\rho^+\rho^-}&=&4g^2_{\rho^+D^+D^0}\frac{(q_{D^+})_\mu
(q_{D^-})_\nu\rho^+_\mu\rho^-_\nu}{m^2_{D^0}-t}\exp[\lambda_D(t-m^2_{D^0})],\nonumber\\
M^{(D^\ast)}_{D^+D^-\to\rho^+\rho^-}&=&g^2_{\rho^+D^+D^{\ast0}}
\varepsilon_{\mu\nu\lambda\sigma}\varepsilon_{\mu^\prime\nu\lambda^\prime\sigma^\prime}
k_\mu k_{\mu^\prime}(q_{D^+})_\lambda
(q_{D^-})_{\lambda^\prime}\rho^+_\sigma\rho^-_{\sigma^\prime}\times\nonumber\\
&&\frac{\exp[\lambda_{D^\ast}(t-m^2_{D^{\ast0}})]}{m^2_{D^{\ast0}}-t},\label{mddvv}\end{eqnarray}
where $k=q_{D^+}-q_{\rho^+}$, $t=k^2$, and $\rho^\pm_\sigma$
stands for the polarization four-vector of the final $\rho^\pm$
meson. The expressions for the amplitude of the reaction $D^0\bar
D^0\to\rho^+\rho^-$ are obtained from Eq.~(\ref{mddvv}) by evident
replacements. When finding the contributions to the imaginary part
of the $\psi(3770)\to\rho^+\rho^-$ decay amplitude, one should
express the polarization four-vectors of moving $\rho^\pm$ mesons
through the polarization three-vectors ${\bm\xi}^{\rho^\pm}$ in
their respective rest frames:
\begin{equation}
(\rho^\pm_0,{\bm\rho}^\pm)=\left(\frac{{(\bm
q}^\pm{\bm\xi}^{\rho^\pm})}{m_\rho}\mbox{,
}{\bm\xi}^{\rho^\pm}+\frac{{\bm q}^\pm({\bm
q}^\pm{\bm\xi}^{\rho^\pm})}{m_\rho(q^\pm_0+m_\rho)}\right),
\label{polar}\end{equation} where $(q_{\rho^\pm})_\mu\equiv
q^\pm_\mu=(q_0^\pm,{\bf q}^\pm)$. Then the $V\to\rho^+\rho^-$
decay amplitude Eq.~(\ref{mvv}) in the rest frame of the decaying
vector meson $V$ can be written as
\begin{eqnarray}
M_{V\to\rho^+\rho^-}=g_1({\bm\xi}{\bm
q}^+)({\bm\xi}^{\rho^+}{\bm\xi}^{\rho^-})+g_2\frac{s^{1/2}}{m_\rho}({\bm\xi}^-{\bm
q}^+)({\bm\xi}^{\rho^+}{\bm\xi})-g_3\frac{s^{1/2}}{m_\rho}({\bm\xi}^+{\bm
q}^+)({\bm\xi}^{\rho^-}{\bm\xi})+\nonumber\\G({\bm\xi}{\bm
q}^+)({\bm\xi}^+{\bm q}^+)({\bm\xi}^-{\bm q}^+),
\label{mvvvnr}\end{eqnarray} where ${\bm\xi}$ is the polarization
three-vector of the decaying $V$, and $G$ is given by
Eq.~(\ref{G}). The results of integration over intermediate state
$D^+D^-$ can be represented as the sum of contributions of the $D$
and $D^\ast$ exchanges
$g_{1,2,3}=g_{1,2,3}^{(D)}+g_{1,2,3}^{(D^\ast)}$,
$G=G^{(D)}+G^{(D^\ast)}$, where
\begin{eqnarray}
g^{(D)}_1&=&N_DA_6,\nonumber\\
g^{(D)}_2&=&-g^{(D)}_3=-\frac{N_D}{s^{1/2}(s^{1/2}+2m_\rho)}\left[A_2+m_\rho(s^{1/2}+m_\rho)A_5-
m_\rho(s^{1/2}+2m_\rho)A_6\right],\nonumber\\
G^{(D)}&=&\frac{N_D}{[m_\rho(s^{1/2}+2m_\rho)]^2}\left\{A_3+(2A_1+A_4)m_\rho(s^{1/2}+m_\rho)
-\frac{2m_\rho(s^{1/2}+2m_\rho)}{({\bm
q}^+)^2}\times\right.\nonumber\\&&\left.\right[B_2+B_5m_\rho(s^{1/2}+m_\rho)
+\frac{1}{2}B_6m_\rho(s^{1/2}+2m_\rho)\left]\right\},\nonumber\\
g^{(D^\ast)}_1&=&N_{D^\ast}\left\{-\left[m^2_{D^\ast}\left(\frac{s}{2}-m^2_\rho\right)+
\frac{1}{4}\left(m^2_{D^\ast}+m^2_\rho-m^2_D\right)\right]A_4+\frac{1}{2}\left(
s-m^2_\rho+m^2_{D^\ast}-\right.\right.\nonumber\\&&\left.\left.m^2_D\right)A_1
-\frac{1}{4}A_3+\left(\frac{s}{2}-m^2_\rho\right)A_6\right\},\nonumber\\
g^{(D^\ast)}_2&=&-g^{(D^\ast)}_3=\frac{m_\rho
N_{D^\ast}}{s^{1/2}}\left\{\left[\left(\frac{s}{2}-m^2_\rho\right)
\frac{s^{1/2}+m_\rho}{s^{1/2}+2m_\rho}+\frac{s^{1/2}}{2m_\rho}
(m^2_{D^\ast}+m^2_\rho-m^2_D)\right]A_5-\right.\nonumber\\&&\left.
\frac{s^{1/2}+m_\rho}{s^{1/2}+2m_\rho}A_2-\left(\frac{s}{2}-m^2_\rho\right)A_6
\right\},\nonumber\\G^{(D^\ast)}&=&N_{D^\ast}\left\{A_4\left[2m^2_{D^\ast}-
\frac{s^{1/2}(s^{1/2}+m_\rho)}{m_\rho(s^{1/2}+2m_\rho)}(m^2_{D^\ast}+m^2_\rho-m^2_D)
-\left(\frac{s}{2}-m^2_\rho\right)\times\right.\right.\nonumber\\&&\left.\left.
\left(\frac{s^{1/2}+m_\rho}{s^{1/2}+2m_\rho}\right)^2-\frac{1}{2m^2_\rho}
(m^2_{D^\ast}+m^2_\rho-m^2_D)^2\right]-\frac{A_3}{(s^{1/2}+2m_\rho)^2}+2A_1+
\frac{B_5}{({\bm
q}^+)^2}\times\right.\nonumber\\&&\left.\left[(s-2m^2_\rho)
\frac{s^{1/2}+m_\rho}{s^{1/2}+2m_\rho}+\frac{s^{1/2}}{m_\rho}
(m^2_{D^\ast}+m^2_\rho-m^2_D)\right]-\frac{2B_2}{({\bm
q}^+)^2}\times\frac{s^{1/2}+m_\rho}{s^{1/2}+2m_\rho}-
\right.\nonumber\\&&\left.\left(\frac{s^{1/2}}{2}-m^2_\rho\right)\frac{B_6}{({\bm
q}^+)^2}\right\}.\label{aid}\end{eqnarray} The normalization
factors are
\begin{eqnarray}
N_D&=&8g_{\psi(3770)D\bar D}g^2_{\rho^0D\bar D},\nonumber\\
N_{D^\ast}&=&-g_{\psi(3770)D\bar D}g^2_{\rho^0D^\ast\bar D},
\label{normfac}\end{eqnarray} where the quark model relations
$|g_{\rho^0D^\ast\bar D}|\approx
\frac{1}{2}g_{\omega\rho\pi}\approx7.2$ GeV$^{-1}$,
$|g_{\rho^0D\bar
D}|\approx|g_{\rho^0KK}|=\frac{1}{\sqrt{2}}|g_{\phi
KK}|\approx3.3$ should be used. The coefficients $A_i$, $i=1$...6
and $B_{2,5,6}$ result from the integration over intermediate
state $D^+D^-$ and look as
\begin{eqnarray}
A_1&=&\frac{|{\bm q}_{D^+}|^2}{8\pi s^{1/2}|{\bm
q}^+|}\int_{-1}^1x\exp[2\lambda_D(a_D+x)|{\bm q}_{D^+}||{\bm
q}^+|]dx \approx\frac{\lambda_D|{\bm q}_{D^+}|^3}{6\pi
s^{1/2}}\times\exp
\left(-\frac{1}{2}\lambda_Ds\right),\nonumber\\
A_2&=&\frac{|{\bm q}_{D^+}|^3}{16\pi
s^{1/2}}\int_{-1}^1(1-x^2)\exp[2\lambda_D(a_D+x)|{\bm
q}_{D^+}||{\bm q}^+|]dx \approx\frac{|{\bm q}_{D^+}|^3}{12\pi
s^{1/2}}\times\exp
\left(-\frac{1}{2}\lambda_Ds\right),\nonumber\\
A_3&=&-\frac{|{\bm q}_{D^+}|^3}{4\pi
s^{1/2}}\int_{-1}^1x(a_D+x)\exp[2\lambda_D(a_D+x)|{\bm
q}_{D^+}||{\bm q}^+|]dx \approx-\frac{|{\bm q}_{D^+}|^3}{6\pi
s^{1/2}}\times\nonumber\\&&\exp
\left(-\frac{1}{2}\lambda_Ds\right)\times
\left(1-\frac{1}{2}\lambda_Ds\right),\nonumber\\
A_4&=&-\frac{|{\bm q}_{D^+}|}{16\pi s^{1/2}|{\bm
q}^+|^2}\int_{-1}^1\frac{x}{a_D+x}\exp[2\lambda_D(a_D+x)|{\bm
q}_{D^+}||{\bm q}^+|]dx \approx\frac{2|{\bm q}_{D^+}|^3}{3\pi
s^{5/2}}\times\nonumber\\&&\exp
\left(-\frac{1}{2}\lambda_Ds\right)\times
\left(1+\frac{1}{2}\lambda_Ds\right),\nonumber\\
A_5&=&-\frac{|{\bm q}_{D^+}|^2}{32\pi s^{1/2}|{\bm
q}^+|}\int_{-1}^1\frac{1-x^2}{a_D+x}\exp[2\lambda_D(a_D+x)|{\bm
q}_{D^+}||{\bm q}^+|]dx \approx\frac{|{\bm q}_{D^+}|^3}{6\pi
s^{3/2}}\times\nonumber\\&&\exp
\left(-\frac{1}{2}\lambda_Ds\right),\nonumber\\
A_6&=&-\frac{|{\bm q}_{D^+}|^3}{32\pi s^{1/2}|{\bm
q}^+|^2}\int_{-1}^1\frac{x(1-x^2)}{a_D+x}\exp[2\lambda_D(a_D+x)|{\bm
q}_{D^+}||{\bm q}^+|]dx \approx\frac{2|{\bm q}_{D^+}|^5}{15\pi
s^{5/2}}\times\nonumber\\&&\exp
\left(-\frac{1}{2}\lambda_Ds\right)\times
\left(1+\frac{1}{2}\lambda_Ds\right),\nonumber\\
B_2&=&\frac{|{\bm q}_{D^+}|^3}{16\pi
s^{1/2}}\int_{-1}^1(3x^2-1)\exp[2\lambda_D(a_D+x)|{\bm
q}_{D^+}||{\bm q}^+|]dx \approx\frac{\lambda_D^2|{\bm
q}_{D^+}|^5s^{1/2}}{60\pi}\times\nonumber\\&&\exp
\left(-\frac{1}{2}\lambda_Ds\right),\nonumber\\
B_5&=&-\frac{|{\bm q}_{D^+}|^2}{32\pi s^{1/2}|{\bm
q}^+|}\int_{-1}^1\frac{3x^2-1}{a_D+x}\exp[2\lambda_D(a_D+x)|{\bm
q}_{D^+}||{\bm q}^+|]dx \approx-\frac{4|{\bm q}_{D^+}|^5}{15\pi
s^{7/2}}\times\nonumber\\&&\exp
\left(-\frac{1}{2}\lambda_Ds\right)\times\left[1+\frac{1}{2}\lambda_Ds+
\frac{1}{2}\times\left(\frac{1}{2}\lambda_Ds\right)^2\right],\nonumber\\
B_6&=&-\frac{|{\bm q}_{D^+}|^3}{32\pi s^{1/2}|{\bm
q}^+|^2}\int_{-1}^1\frac{x(5x^2-3)}{a_D+x}\exp[2\lambda_D(a_D+x)|{\bm
q}_{D^+}||{\bm q}^+|]dx \approx\frac{8\lambda_D|{\bm
q}_{D^+}|^7}{35\pi s^{5/2}}\times\nonumber\\&&\exp
\left(-\frac{1}{2}\lambda_Ds\right)\times
\left(1+\frac{1}{4}\lambda_Ds\right). \label{coeff}\end{eqnarray}
Here
\begin{equation}
a_D=\frac{2m^2_\rho-s}{4|{\bm q}_{D^+}||{\bm
q}^+|},\label{ad}\end{equation} and the approximate equality of
the slopes of the $D$ and $D^\ast$ exchanges
$\lambda_{D^\ast}\approx\lambda_D$ is supposed. The approximate
equalities in Eq.~(\ref{coeff}) are valid in the threshold
situation. One should have in mind that $m_{D^\ast}\equiv
m_{D^{\ast0}}$, $m_D\equiv m_{D^+}$ in Eqs.~(\ref{aid}),
(\ref{coeff}). To obtain the contribution of the intermediate
state $D^0\bar D^0$ one should make the replacement $m_{D^\ast}\to
m_{D^{\ast+}}$, $m_D\to m_{D^0}$ in the above expressions. To get
the final expressions, one should take, respectively,  the sum
(difference) of the $D^+D^-$ and $D^0\bar D^0$ contributions in
the case of the conservation (violation) of isospin in the course
of the decay.


\begin{thebibliography}{99}
\bibitem{ach94}
N.~N.~Achasov and A.~A.~Kozhevnikov, Phys. Rev. D {\bf 49}, 275
(1994);\\ Phys. Lett. B {\bf 260}, 425 (1991); Pis'ma v ZhETF {\bf
54}, 197 (1991) [JETP Lett. {\bf54},193 (1991)].
\bibitem{lipkin87}
H.~J.~Lipkin. Nucl. Phys. B {\bf 291}, 720 (1987).
\bibitem{jon}
J.L. Rosner, hep-ph/0405196.
\bibitem{pdg}
S.~I.~Eidelman {\it et al.} Phys. Lett. B {\bf 592}, 1 (2004).
\bibitem{asner}
D.~Asner,  hep-ex/0405009.
\bibitem{babar}
B.~Aubert {\it et al.}, hep-ex/0408022.
\bibitem{belle}
J.~Draqic {\it et al.}, Phys. Rev. Lett. {\bf93}, 131802 (2004).
\bibitem{slopes}
K.~G.~Boreskov and A.~B.~Kaidalov, Yad. Fiz. {\bf37}, 174
(1983) [Sov.J.Nucl.Phys. {\bf 37}, 100 (1983)];\\
 W.~Cassing, L.~A.~Kondratyuk,
G.~I.~Lykasov, M.~V.~Rzjanin,  Phys. Lett. B {\bf 513}, 1 (2001).
\bibitem{rosner}
J.L. Rosner, hep-ph/0411003.
\bibitem{cleo}
CLEO Collaboration, hep-ex/0504003.
\bibitem{babar05}
BABAR Collaboration, hep-ex/0504001.
\bibitem{1969}
E. Cremer and N. Gourdin, Nucl. Phys. {\bf B9}, 451 (1969).
\bibitem{lepage}
G.P. Lepage, Phys. Rev. D {\bf42}, 3251 (1990).
\bibitem{formfactorsinpsi}
N.~N.~Achasov and A.~A.~Kozhevnikov, Phys. Rev. D {\bf 58}, 097502
(1998);\\ Yad. Fiz. {\bf 62}, 364 (1999) [Phys.Atom.Nucl. {\bf62},
328 (1999)].
\end{thebibliography}
\end{document}